\documentclass[twocolumn,showpacs,preprintnumbers,amsmath,amssymb]{revtex4}
\pdfoutput=1

\usepackage{graphicx}
\usepackage{dcolumn}
\usepackage{bm}
\usepackage{multirow} 
\graphicspath{{Raghunathan_anisotropy_figures/}}

\begin{document}

\preprint{}

\title{A Theoretical Approach for Computing Magnetic Anisotropy 
in\\Single Molecule Magnets}

\author{Rajamani Raghunathan$^\$$, S. Ramasesha$^\$$
and Diptiman Sen$^\#$}
 \affiliation{$^\$$Solid State and Structural Chemistry 
Unit, Indian Institute of Science, Bangalore – 560012. INDIA.\\
$^\#$Centre for High Energy Physics, Indian Institute of Science, 
Bangalore – 560012. INDIA.} 
\email{ramasesh@sscu.iisc.ernet.in}

\date{\today}

\begin{abstract}
We present a theoretical approach to calculate the molecular
magnetic anisotropy parameters, $D_M$ and $E_M$ for single molecule
magnets in any eigenstate of the exchange Hamiltonian, treating
the anisotropy Hamiltonian as a perturbation. Neglecting inter-site
dipolar interactions, we calculate molecular magnetic anisotropy in
a given total spin state from the known single-ion anisotropies of
the transition metal centers. The method is applied to $Mn_{12}Ac$
and $Fe_8$ in their ground and first few excited eigenstates, as an
illustration. We have also studied the effect of orientation of
local anisotropies on the molecular anisotropy in various eigenstates
of the exchange Hamiltonian. We find that, in case of $Mn_{12}Ac$,
the molecular anisotropy depends strongly on the orientation of the
local anisotropies and the spin of the state. The $D_M$ value of
$Mn_{12}Ac$ is almost independent of the orientation of the local
anisotropy of the core $Mn(IV)$ ions. In the case of $Fe_8$, the
dependence of molecular anisotropy on the spin of the state in
question is weaker.
\end{abstract}

                             
                              
\maketitle

\section{\label{sec:intro}Introduction \protect }

Following the synthesis and the discovery of exotic properties 
such as quantum resonant tunneling (QRT) in the single molecule 
magnet (SMM) $Mn_{12}Ac$ during the 1990s, there has been a flurry 
of activity in the field of molecular magnetism \cite{thomas1, 
linert1, miller1, shoji1}. This has led to the 
synthesis of new systems such as $Fe_8$ as well as to 
the observation of new phenomena such as quantum coherence 
\cite{wieghardt1, barco1}. SMMs are mainly high nuclearity transition 
metal complexes with a high-spin ground state ($S_{GS}$). They are also 
characterized by large uniaxial magnetic anisotropy \cite{sessoli1}. 
The Hamiltonian corresponding to the magnetic anisotropy of a molecular 
system can be written as,
\begin{eqnarray}
\label{eqn1}
\hat{H}_D=\hat{S}_M \cdot \mathcal{D}^{(M)} \cdot \hat{S}_M
\end{eqnarray}
where, $\hat{S}_M$ is the spin operator for the total spin of the molecule 
and $\mathcal{D}^{(M)}$ is the magnetic anisotropy tensor of the molecule. 
In usual practice, the anisotropy tensor is diagonalized and the 
principle axis of the molecule would correspond to the eigenvectors of 
the tensor. Since in most physical situations, the quantity of interest 
is the energy gaps between the otherwise degenerate states split by 
the magnetic anisotropy, the condition of zero trace is imposed on 
the $\mathcal{D}^{(M)}$ tensor. If, $D_{XX}^M$, $D_{YY}^M$ and $D_{ZZ}^M$ 
are the molecular anisotropies along the three principal directions such 
that $D_{XX}^M+D_{YY}^M+D_{ZZ}^M=0$, we can define two parameters, 
$D_M$ and $E_M$ given by, 
\begin{eqnarray}
D_M&=&D_{ZZ}^M-\frac{1}{2}\left(D_{XX}^M+D_{YY}^M \right) \nonumber \\
E_M&=&\frac{1}{2}\left(D_{XX}^M-D_{YY}^M \right) \nonumber
\end{eqnarray}
where, $D_M$ and $E_M$ are called the axial and rhombic anisotropies 
respectively. This leads to the common form of the magnetic anisotropy 
Hamiltonian of a SMM, 
\begin{eqnarray}
\label{eqn2}
\hat{H}_M=D_M\left(\hat{S}_Z^2-\frac{1}{3}S(S+1)\right)+
E_M\left(\hat{S}_X^2-\hat{S}_Y^2\right)
\end{eqnarray}
For the single molecule magnet to have nonzero magnetization in 
the ground state, it is necessary that the anisotropy constant $D_M$ 
in the spin Hamiltonian (Eq. \ref{eqn2}) of the complex be negative; this 
ensures that the ground state of the system then would correspond 
to the highest magnetization state of the molecule, in its high-spin 
ground state. This requirement of negative ‘$D_M$’, besides a 
high-spin ground state makes it hard to tailor SMMs. The second order 
transverse or rhombic anisotropy given by the last term in Eq. \ref{eqn2}, 
allows transition between states with spin $S$ that differ in their 
$M_s$ values by two. $E_M$ will be zero if the $S_X^2-S_Y^2$ operator 
does not remain invariant under symmetry of the molecule. In this case 
higher order spin-spin interaction terms are required to observe QRT. 
For example, the $D_{2d}$ symmetry in $Mn_{12}Ac$ prohibits the 
existence of first order rhombic anisotropy. Thus, the parameters 
$D_M$ and $E_M$ govern the quantum tunneling properties of a SMM 
and inputs from theoretical modeling could help in designing the 
architecture for the synthesis of SMMs. 

Theoretical modeling of SMMs presents two difficulties. Firstly, 
the complexes contain many spin centers, and often these centers 
have different spins as in the case of $Mn_{12}Ac$, where the 
four $Mn(IV)$ ions have spin-3/2, while the eight $Mn(III)$ ions 
have spin-2. In these systems, usually there exist multiple exchange 
pathways between any given pair of ions leading to uncertain 
magnitude and sign of the magnetic interactions in the system. 
The topology of magnetic interactions also is often such as to 
result in magnetic frustration leading to closely spaced low-lying 
states of unpredictable total spin. Thus, solving even the simple 
Heisenberg exchange Hamiltonian of these systems turns out to be a 
challenge. If we do not employ a spin adapted basis to set-up the 
Hamiltonian matrix we could encounter convergence difficulties 
associated with closely spaced eigenvalues even when the corresponding 
eigenstates belong to different total spin sectors. However, the 
assorted spin cluster that a SMM is renders construction of spin 
adapted basis difficult. This problem has been addressed by resorting 
to a valence bond scheme for construction of the spin adapted basis 
and we are now in a position to block-diagonalize the Hamiltonian by 
exploiting both spin and spatial symmetries \cite{sahoo1}. Thus, while the 
challenge of accurately solving the exchange Hamiltonian of a large 
assorted spin system with arbitrary topology of exchange interactions 
is within grasp, the challenge of computing the magnetic anisotropy 
constants of a SMM still remains.

\begin{figure}
\includegraphics[height=6cm]{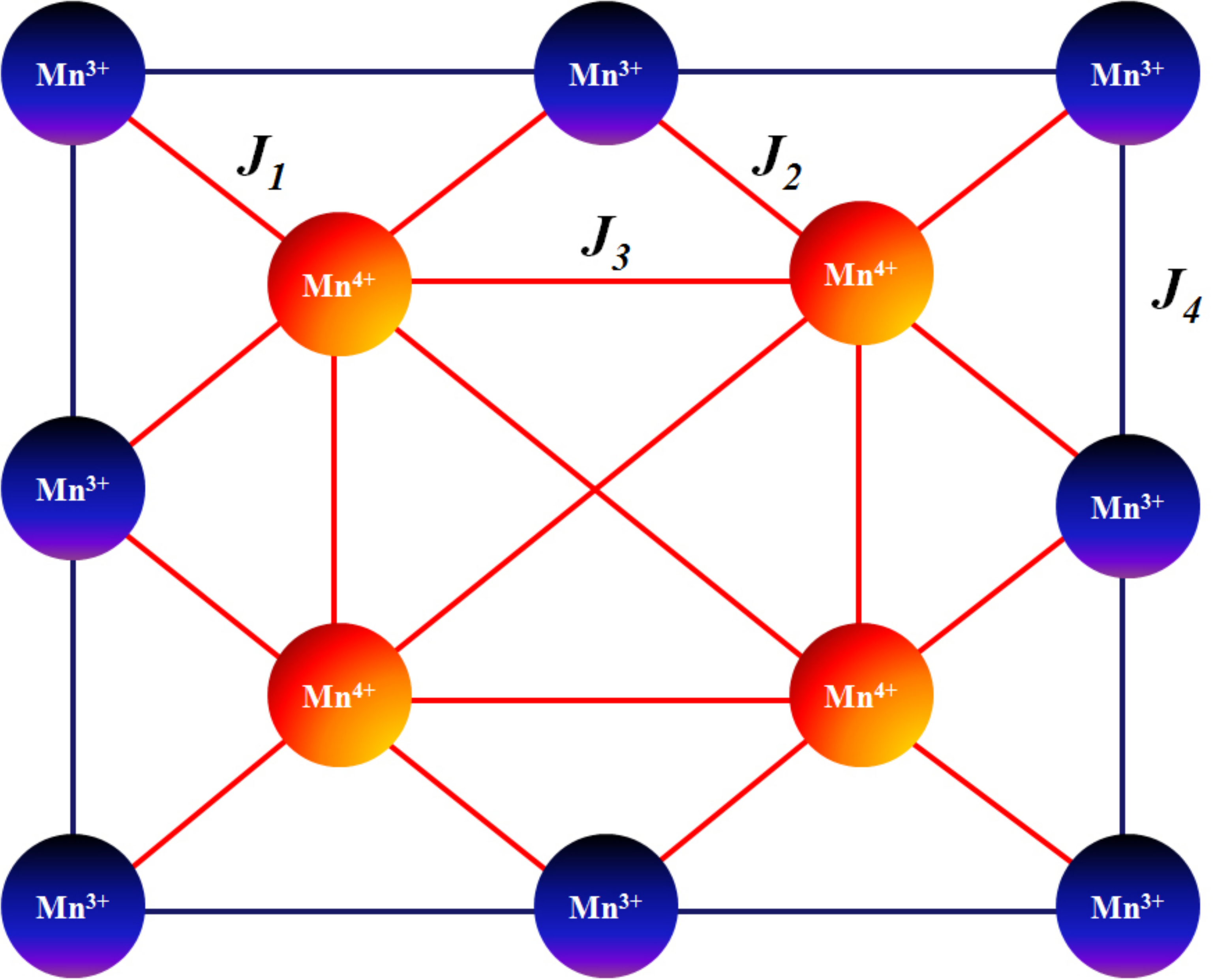}
\caption{\label{mn12schematic} Schematic of possible exchange 
interactions in $Mn_{12}Ac$ SMM. The peripheral $Mn(III)$ ions 
represented by blue circles correspond to spin-2 sites and those 
represented by yellow circles are the core $Mn(IV)$ ions each of 
spin-3/2. J’s are the strength of superexchange interaction with 
$J_1$=215 K, $J_2$=$J_3$=85.6 K, $J_4$=-64.5 K \cite{raghu1}.}
\end{figure}

The magnetic anisotropy in an isolated ion arises from explicit 
dipolar interactions between the unpaired electrons in the magnetic 
center as well as from relativistic (spin-orbit) interactions. 
The former is the main origin of zero field splitting observed in 
triplet states of conjugated organic molecules such as naphthalene 
\cite{sinnecker1}. However, in systems containing heavier elements 
the relativistic effects dominate. In a system with several magnetic 
centers such as a SMM, given the spin and single ion anisotropy of 
each magnetic center, the magnetic anisotropy could arise both from 
dipolar interactions between magnetic centers and relativistic effects. 
Theoretically, magnetic anisotropy of an isolated magnetic center 
in an appropriate ligand environment can be computed within density 
functional techniques (DFT) \cite{baruah1,arino1}. However, in the 
case of the SMMs with many magnetic centers, such a calculation is both 
conceptually and computationally difficult, since DFT methods do 
not conserve total spin. The usual approach in these cases is to 
carry out simple tensoral summation of the anisotropies of the 
constituent magnetic centers to obtain the magnetic anisotropy in 
the SMMs along the lines of an oriented gas model employed in the 
calculation of macroscopic nonlinear optic (NLO) coefficients from 
isolated molecular NLO coefficients \cite{oudar1,bencini1,bencini2}. 
Such an approach, in the case of SMMs, suffers from the 
drawback that the anisotropy constants so computed are 
independent of the total spin state of the molecule. 
In this paper, we treat the anisotropic magnetic interaction between 
the magnetic centers as a perturbation over the magnetic exchange 
Hamiltonian of the SMM. We exactly solve for the various total spin 
states of the unperturbed Hamiltonian and from the desired eigenstates 
obtain the spin-spin correlation functions necessary to compute the 
anisotropy constants. In the next section we describe the method in 
detail. In the third section we present the results of our studies on 
the two SMMs, $Mn_{12}Ac$ and $Fe_8$. In the fourth section we summarize 
our studies and outline the extension of this method for the computation of 
higher order magnetic anisotropy constants.  

\begin{figure}
\includegraphics[height=6cm]{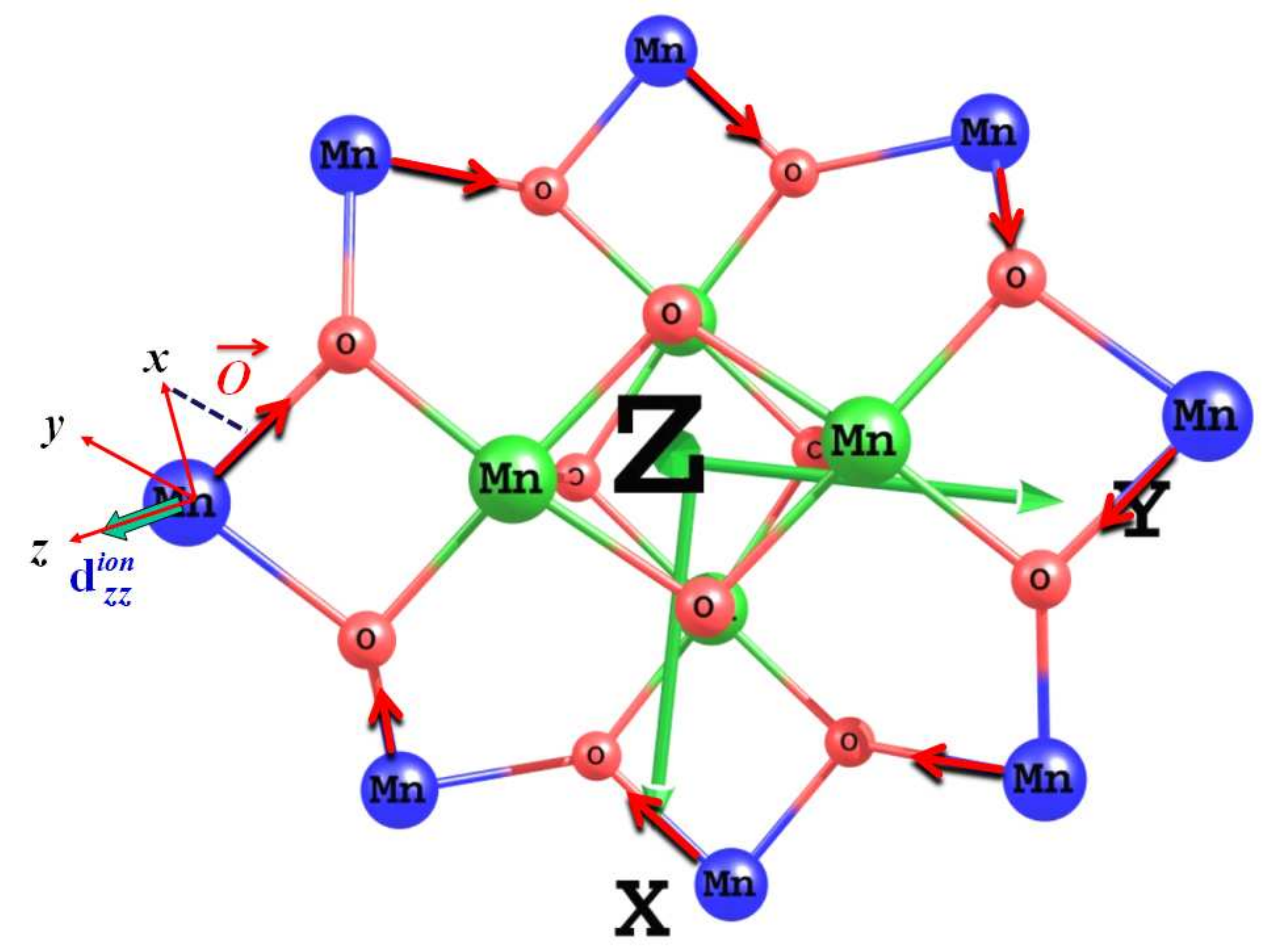}
\caption{\label{mn12localaxis} Schematic of local ($x$, $y$, $z$) 
and laboratory ($X$, $Y$, $Z$) coordinate axes in $Mn_{12}Ac$. 
The blue, green and red spheres correspond to $Mn(III)$ 
(spin-2), $Mn(IV)$ (spin-3/2) and oxygen ions respectively. The 
arrows indicate the $Mn$-$O$ bonds on which the chosen local $x$-axis 
has maximum projection.}
\end{figure}

\section{\label{sec:method}Formulation of the Method}
We treat the exchange Hamiltonian between magnetic centers in the 
SMMs as the unperturbed Hamiltonian,
\begin{eqnarray}
\label{eqn3}
\hat{H}_0=\sum_{\langle ij \rangle} J_{ij} \hat{S}_i \cdot \hat{S}_j
\end{eqnarray}

where, $\langle ij \rangle$ runs over all pairs of centers in the 
model for which the exchange constant is nonzero, $\hat{S}_i$ is 
the spin on the {\it i}th magnetic center. In SMMs such as $Mn_{12}Ac$ 
the spins at all the magnetic centers are not the same and the 
exchange interactions are shown in Fig. \ref{mn12schematic}. 
$H_0$ can be solved exactly for a few low-lying states in a chosen 
spin sector by using methods that have been described in detail 
elsewhere \cite{raghu1}.  

\begin{figure}
\includegraphics[height=4.5cm]{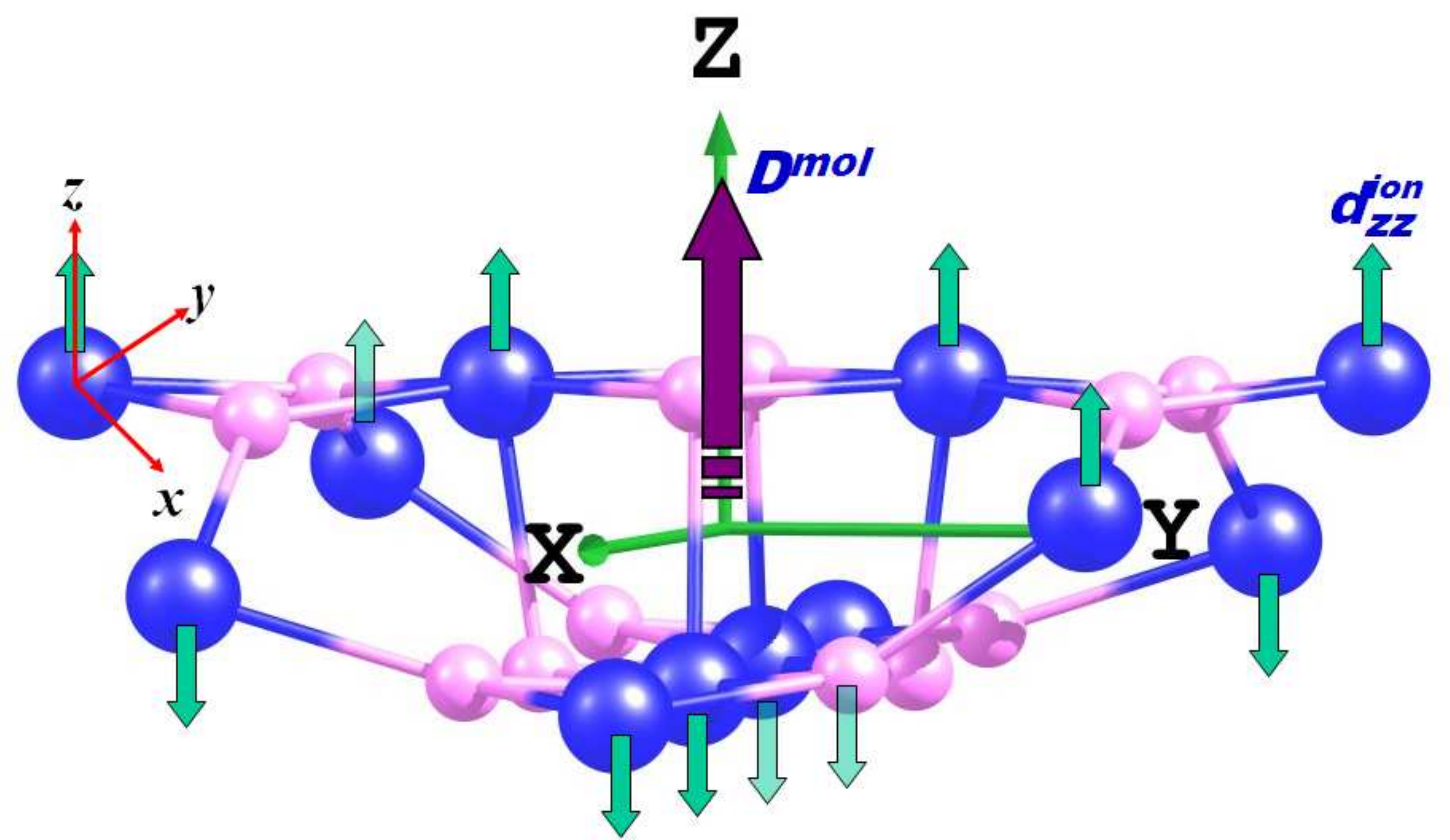}
\caption{\label{mn12scheme1} Schematic diagram showing the 
directions of local anisotropy in $Mn_{12}Ac$. The single-ion 
anisotropies of all the $Mn$ ions are directed along the laboratory 
$Z$ axis (Scheme 1).}
\end{figure}

The general anisotropic interactions in a collection of magnetic 
centers is treated as a perturbation with Hamiltonian $\hat{H}_1^\prime$ 
is given by,
\begin{eqnarray}
\label{eqn4}
\hat{H}_1^\prime=\frac{1}{2}\sum_i\sum_j\sum_\alpha\sum_\beta 
D_{ij,\alpha\beta}\hat{S}_i^\alpha\hat{S}_j^\beta
\end{eqnarray}
where, the indices {\it i} and {\it j} run over all the magnetic 
centers and $\alpha$ and $\beta$ run over $x$, $y$ and $z$ directions 
of the ion. The contributions to inter-center anisotropy constant 
arise due to dipolar interaction between the spins on the two 
centers as well as due to relativistic effects. In the former, 
$D_{ij,\alpha\beta}$ is given by,
\begin{eqnarray}
\label{eqn5}
D_{ij,\alpha\beta}=\frac{1}{2}g^2\mu_B^2 
\left< \frac{\mathcal{R}_{ij}^2\delta_{\alpha\beta}-3R_{ij,\alpha}R_{ij,\beta}}{R_{ij}^5} \right>
\end{eqnarray}
where $\mathcal{R}_{ij}$ ($R_{ij}$) is the vector (distance) between 
the magnetic centers ‘{\it i}’ and ‘{\it j}’, {\it g} is the 
gyromagnetic ratio and $\mu_B$ is the electronic Bohr magneton; the 
expectation value in Eq. \ref{eqn5} is obtained by integration over spatial 
coordinates \cite{carrington1}. Approximating the expectation values of 
the distances by the equilibrium distances, the $D_{ij,\alpha\beta}$ in 
Eq. \ref{eqn5} and by computing the necessary spin-spin correlation functions, 
we can obtain the molecular $\mathcal{D}^{(M)}_{\alpha\beta}$ 
tensor \cite{ramasesha1}. The eigenvalues of this matrix give 
the principal anisotropy values and imposing the condition 
of zero trace of the matrix yields molecular magnetic 
anisotropy constants due to spin-spin interactions. Our computation 
of the magnetic anisotropy constants, assuming only spin 
dipolar interactions for the SMMs $Mn_{12}Ac$ and $Fe_8$ gives 
negligible values of the anisotropy constants compared to the 
experimental values of $D_M$ = -0.7 K and -0.28 K respectively 
in the S=10 ground state \cite{barra1,barra2}. Hence in what follows, 
we completely neglect the contribution of spin-dipolar interactions and 
focus only on the magnetic anisotropy of the SMMs arising from the 
anisotropy of individual magnetic centers. The latter is a 
consequence of mainly spin-orbit interactions. We now assume that 
the interactions responsible for magnetic anisotropy are short 
ranged and neglect inter-center contributions to magnetic anisotropy 
in Eq. \ref{eqn4}. This is justified since relativistic (spin-orbit) 
interactions, largely responsible for the anisotropy is short 
ranged (falling off as $1/r^3$) and the distances between the magnetic 
centers is much larger compared to the ionic radius of the transition 
metal ion. The resulting perturbation term is given by,

\begin{figure}
\includegraphics[height=4.5cm]{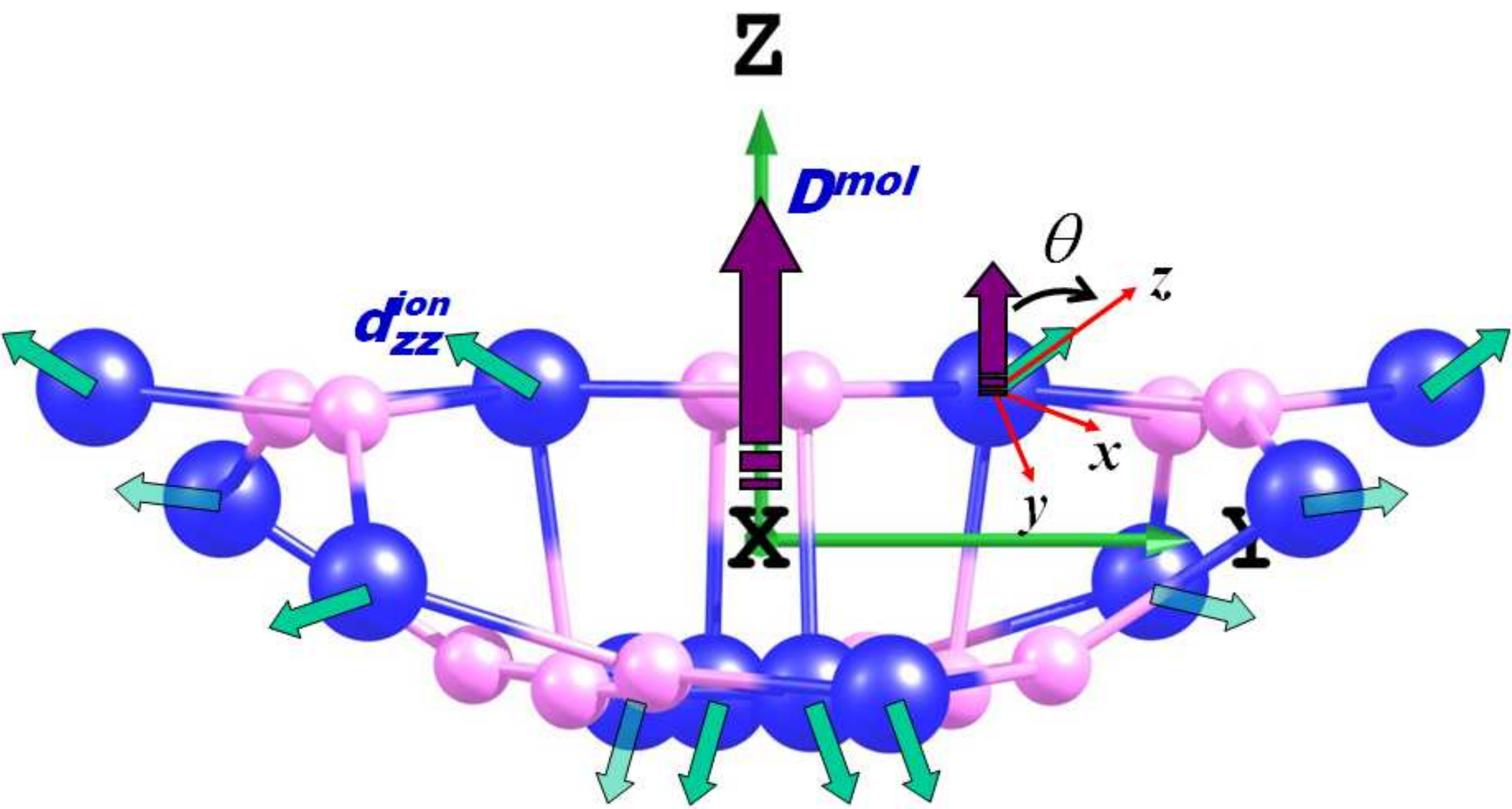}
\caption{\label{mn12scheme2} Schematic diagram showing the 
directions of local anisotropy in $Mn_{12}Ac$. The $z$-component 
of the single-ion anisotropies of all the $Mn(III)$ ions are 
inclined at an angle $\theta$ to the laboratory $Z$ and while 
that of the $Mn(IV)$ ions are kept fixed at $\sim 48^\circ$ 
(Scheme 2).}
\end{figure}

\begin{eqnarray}
\label{eqn6}
\hat{H}_1=\sum_i\sum_{\alpha}\sum_{\beta}D_{i,\alpha\beta}
\hat{S}_i^\alpha\hat{S}_i^\beta
\end{eqnarray}
where, only on-site terms are retained. If the individual magnetic 
centers have different principal axes then we choose a laboratory 
frame and project the local tensor components on to the laboratory 
frame. In such a case, Eq. \ref{eqn6} gets modified to,
\begin{eqnarray}
\label{eqn7}
\hat{H}_1=\sum_i\sum_{\alpha}\sum_{\beta}\sum_l\sum_m
C_{i,l\alpha} C_{i,m\beta} D_{i,\alpha\beta} \hat{S}_i^\alpha\hat{S}_i^\beta
\end{eqnarray}
Where, $C_{i,l\alpha}$ are the direction cosines of the local axis 
of the {\it i}th magnetic center with the {\it l}({\it m}) being the 
coordinate of the laboratory frame and $\alpha$($\beta$) being the 
local coordinates. Since, the Hamiltonians in Eq. \ref{eqn1} and \ref{eqn7}
are equivalent, we can equate the matrix elements 
$\langle n,S_M,M|\hat{H}_1|n,S_M,M^\prime \rangle$ and 
$\langle S_M,M|\hat{H}_M|S_M,M^\prime \rangle$ for any pair of 
eigenstates of the exchange Hamiltonian in Eq. \ref{eqn3}; $|M \rangle$ and 
$|M^\prime \rangle$ correspond to a state ‘{\it n}’ with spin $S_M$ 
in which we are interested. Calculating these matrix elements for 
$\hat{H}_M$ is straightforward from the algebra of spin operators. 
However, evaluation of these matrix elements between eigenstates of 
$\hat{H}_0$, requires computing them in the basis of the spin 
orientation of the sites. From a given eigenstate of $\hat{H}_0$, 
$|n,S,M \rangle$, we can compute all eigenstates with different $M$ 
values by using ladder operators corresponding to spin $S$. 
Given a $S$ value, the above condition would give rise to $(2S+1)^2$ 
equations, while the tensor $\mathcal{D}^{(M)}$, has only nine 
components. Thus, for the $Mn_{12}Ac$ system, with ground state 
spin of 10, there would be 441 equations and we have more equations 
than unknowns. However, we could take any nine equations and solve 
for the components of the tensor $\mathcal{D}^{(M)}$ and we would get 
unique values of the components. This is guaranteed by the 
Wigner-Eckart theorem and we have also verified this by solving for 
the $\mathcal{D}^{(M)}$ tensor from several arbitrarily different 
selections of the nine equations.

\begin{figure}
\includegraphics[height=6cm]{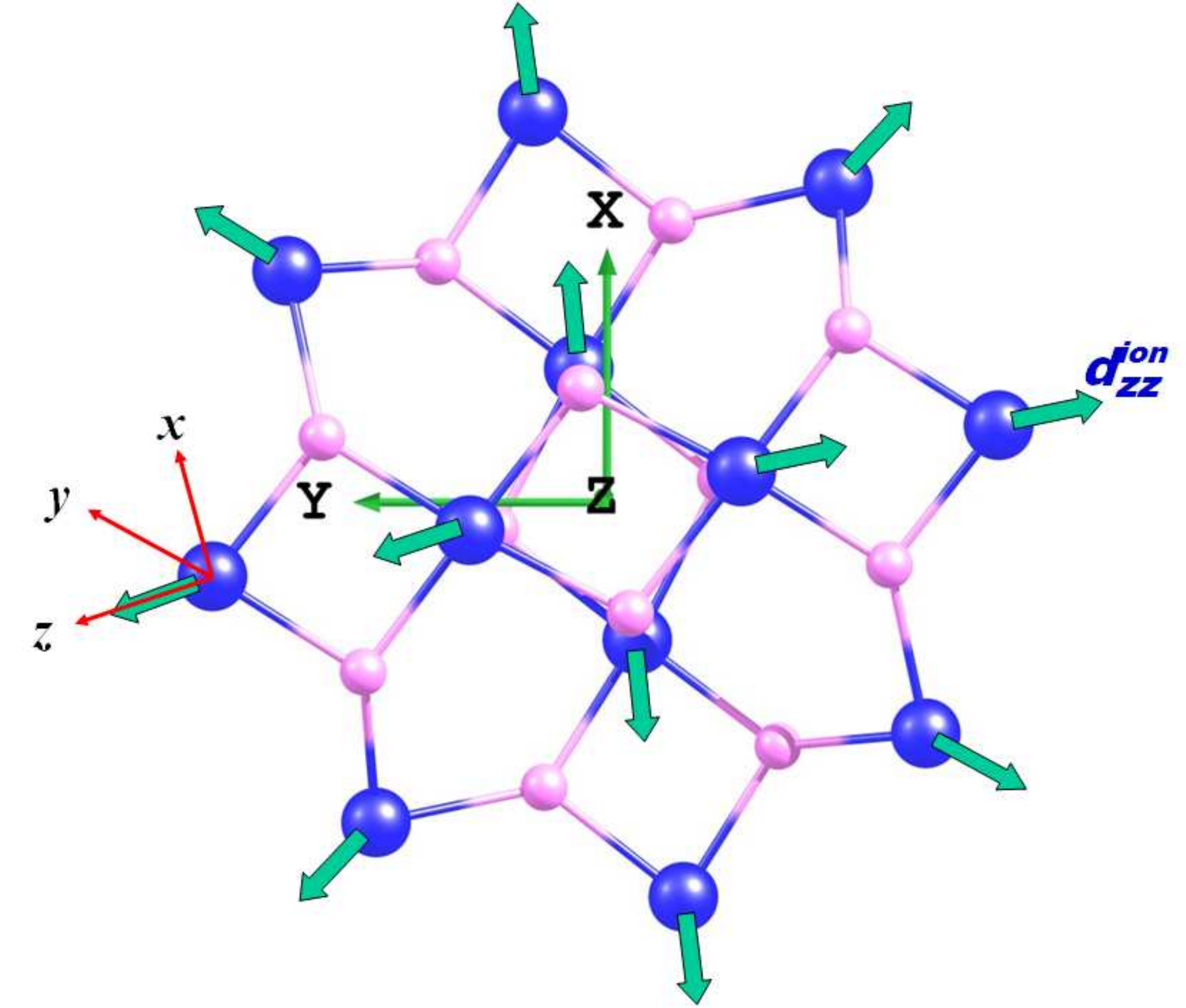}
\caption{\label{mn12scheme3} Schematic diagram showing the 
directions of local anisotropy in $Mn_{12}Ac$. The $z$-component 
of the single-ion anisotropies of all the $Mn$ ions are directed 
along the plane perpendicular to the laboratory $Z$ axis (Scheme 3).}
\end{figure}

\section{\label{sec:results} Results and Discussion}
\begin{figure}
\includegraphics[height=4.5cm]{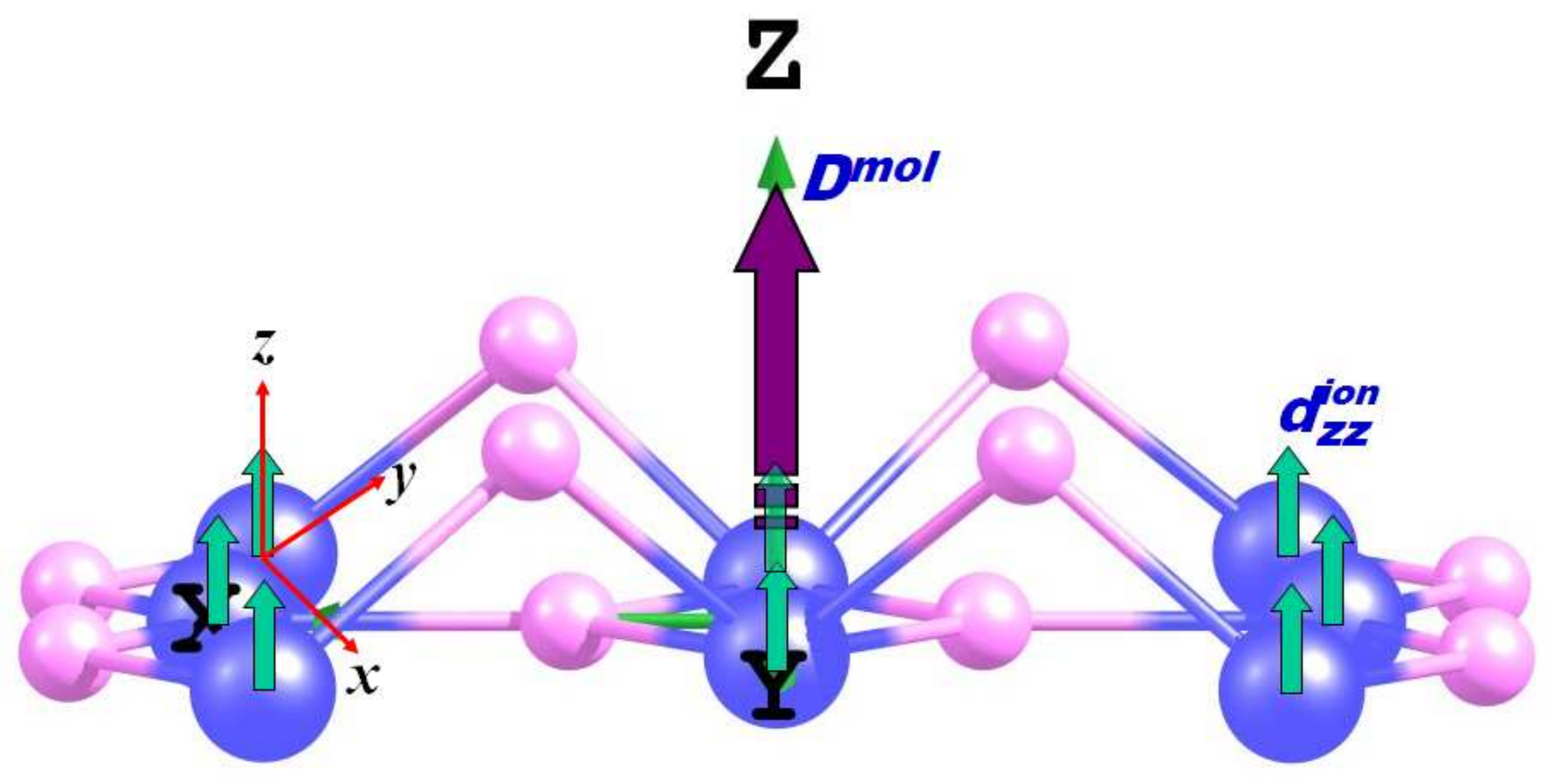}
\caption{\label{fe8scheme1} Schematic diagram showing the 
directions of local anisotropy in $Fe_8$. The single-ion anisotropies 
of all the $Fe(III)$ ions are directed along the laboratory $Z$ axis 
(Scheme 4).}
\end{figure}

We have computed $D_M$ and $E_M$ values for both $Mn_{12}Ac$ and 
$Fe_8$ systems for different orientations of local anisotropy. 
We have used the single-ion anisotropy values quoted in the 
literature for complexes of these ions in similar ligand 
environments \cite{cornia1,barra3,accorsi1,heerdt1}. 
While discussing the results, we refer to 
the local axis of the ions as $x$, $y$ and $z$ and the 
laboratory axis is denoted as $X$, $Y$ and $Z$. The laboratory 
frame we choose can be arbitrary. This is because, on determining 
$\mathcal{D}^{(M)}$ in the laboratory frame, we diagonalize it and 
the principal axis of the molecule is given by the eigenvectors of 
the $\mathcal{D}^{(M)}$ matrix. The principal axes of the molecule are 
unique and do not depend on the laboratory frame that is selected. 
We have computed the anisotropy parameters for both these systems as a 
function of the angle $\theta$ which the $z$-axis of the ion makes with 
the laboratory $Z$-axis. The orientation of $z$-component of the 
single-ion anisotropy in every site is shown in Fig. \ref{mn12scheme1}, 
\ref{mn12scheme2} and \ref{mn12scheme3} (schemes 1, 2 and 3) 
for $Mn_{12}Ac$ and in Fig. \ref{fe8scheme1}, \ref{fe8scheme2}, 
and \ref{fe8scheme3} (schemes 4, 5 and 6) for $Fe_8$ systems 
respectively. Once the $z$-axis ($\vec{z}$) of the ion is fixed, 
then $\vec{x}$ is obtained by Gram-Schmidt orthogonalization procedure. 
Though the choice of this vector is arbitrary in a plane perpendicular 
to z-axis, we have fixed the  direction of $\vec{x}$ such as to 
have maximum projection along a $M$-$O$ ($M$=$Mn$, $Fe$) bond in $Mn_{12}Ac$ 
 as well as in $Fe_8$ (Fig. \ref{mn12localaxis} and Fig. \ref{fe8scheme3}). 
If $\vec{O}$  is the vector connecting a $M$ site and 
a neighbouring $O$ ion, then we obtain $\vec{x}$ from,
\begin{eqnarray}
\label{eqn8}
\vec{x}=\vec{O}-\left(\vec{O} \cdot \vec{z}\right) \vec{z}
\end{eqnarray}
Then, the y-axis of the ion is obtained by taking cross product 
of the $\vec{z}$ and $\vec{x}$, $\vec{y}=\vec{z}\times\vec{x}$. 
These three mutually orthogonal vectors are then normalized to 
obtain the orthonormal set of coordinate axes $x$, $y$ and $z$ 
of the ion centre. The single ion local axes is represented in 
the laboratory frame as, 
\begin{eqnarray}
\label{eqn9}
x&=&C_{i,Xx}X+C_{i,Yx}Y+C_{i,Zx}Z\nonumber\\
y&=&C_{i,Xy}X+C_{i,Yy}Y+C_{i,Zy}Z\nonumber\\
z&=&C_{i,Xz}X+C_{i,Yz}Y+C_{i,Zz}Z
\end{eqnarray}
where, C’s are the direction cosines of Eq. \ref{eqn7} and the index {\it i} 
correspond to the site {\it i}. The procedure is repeated for every 
magnetic ion to obtain coordinate axes set $x$, $y$ and $z$ and the 
direction cosines in each case. We have obtained the magnetic 
anisotropy parameters $D_M$ and $E_M$ for $Mn_{12}Ac$ and $Fe_8$ 
clusters as a function of the angle of rotation of the local $z$-axis 
with respect to the laboratory $Z$-axis. In the following subsections, 
we discuss the results for the two clusters.

\begin{figure}
\includegraphics[height=7cm]{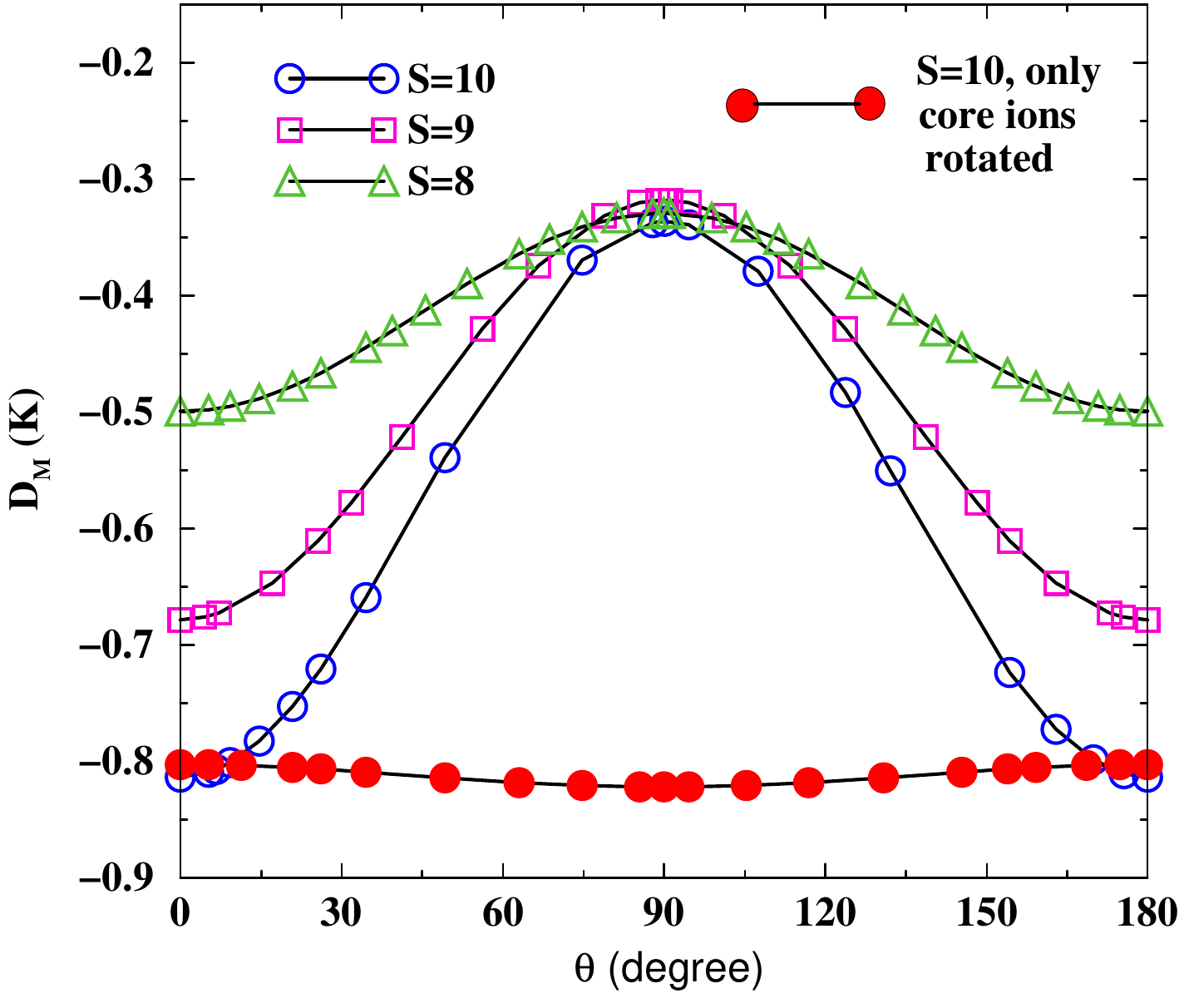}
\caption{\label{mn12Dplot} Variation of $D_M$ as a 
function of $\theta$, the angle the $z$-component of local 
anisotropy of $Mn(III)$ ions makes with the laboratory $Z$-axis 
for scheme 2, in eigenstates with total spin 10, 9 and 8. 
The orientation of $Mn(IV)$ ions is kept fixed at $\sim 48^\circ$ 
from the molecular $Z$-axis. The curve with filled circles  
correspond to the variation of $D_M$, when the local anisotropies 
of the core $Mn(IV)$ ions only are rotated and those of $Mn(III)$ 
ions are fixed along the $Z$ axis.}
\end{figure}

\begin{table}
\caption{\label{table1} $D_M$ values of ground and excited 
states of $Mn_{12}Ac$ under various schemes in K. For scheme 2, 
we have presented the $D_M$ values only for $\theta$=26.2$^\circ$ 
for which the $D_M$ value of the ground state matches with the 
experimentally observed value.}
~~\\
\begin{tabular}{|c|c|c|c|}\hline
\multirow{2}{*}{\bf State} & \multicolumn{3}{c|}{$\mathbf{D_M}$ {\bf (K)}}\\\cline{2-4}
      & {\bf Scheme 1}  & {\bf Scheme 2}  & {\bf Scheme 3}\\\hline
Ground state (S=10) & -0.8028 & -0.7209 & 0.3991 \\\hline
First excited state & \multirow{2}{*} {-0.6722} & \multirow{2}{*} {-0.6105} & \multirow{2}{*} {0.3341} \\
(S=9) $E_g$=35.1 K &  &  &    \\\hline 
Second excited state & \multirow{2}{*} {-0.5009} & \multirow{2}{*} {-0.4664} & \multirow{2}{*} {0.2488} \\ 
(S=8) $E_g$=60.4 K &  &  &    \\\hline
\end{tabular}
\end{table}

\subsection{\label{sec:Mn12Ac_res}Magnetic anisotropy in $Mn_{12}Ac$ SMM}
We have first obtained the ground state and few excited states 
of the $Mn_{12}Ac$ system by exactly solving the unperturbed Hamiltonian 
given in Eq. \ref{eqn3}, using the exchange interactions shown in 
Fig. \ref{mn12schematic} \cite{raghu1}. The ground state of 
the system corresponds to total spin 10 with a total spin 9 
excited state at 35.1 K from the ground state. The second excited 
state occurs at 60.4 K from the ground state and corresponds to 
total spin 8. To obtain the molecular anisotropy values in these 
eigenstates, we have used different single-ion axial anisotropy 
values of -5.35 K and 1.226 K respectively for $Mn(III)$ and $Mn(IV)$ 
ions. We have also introduced transverse anisotropy of 0.022 K and 
0.043 K for $Mn(III)$ and $Mn(IV)$ sites respectively. We have 
studied the variation of molecular anisotropy as a function of 
orientation of the local anisotropies by rotating the local 
$\mathcal{D}$ tensor around the molecular 
$Z$-axis. Scheme 1 shown in Fig. \ref{mn12scheme1}
corresponds to the case when all the single-ion $z$ axes are pointed 
parallel to the laboratory $Z$ direction. In scheme 2 (Fig. \ref{mn12scheme3}), 
we have fixed the orientation of the local anisotropies of the 
core $Mn(IV)$ ions along the line joining the ion and the molecular 
centre ($\sim$ 48$^\circ$ from the laboratory $Z$-axis), while 
the anisotropies of the $Mn(III)$ ions is rotated and the angle 
which it makes with the molecular $Z$-axis is defined as $\theta$ 
(refer Fig. \ref{mn12scheme2}). The orientation of the 
local anisotropy of $Mn(III)$ ions for which we get the 
best agreement with experiments corresponds to $\theta=26^\circ$. 
In scheme 3 (Fig. \ref{mn12scheme3}), we have restricted the $z$-component 
of single-ion anisotropy to the laboratory $X-Y$ plane. We have studied 
the variation in the molecular anisotropies in these schemes for 
ground and the excited eigenstates. We show in Table \ref{table1}, 
the $D_M$ values for the ground and the excited spin states of the 
molecule for schemes 1, 2 and 3. We note that when the local 
anisotropies are systematically varied, there is a very large 
variation in the molecular anisotropy as a function of the local 
orientation (Fig. \ref{mn12Dplot}). This seems to be true for all 
the states of $Mn_{12}Ac$ we have studied. We note that given the 
orientations of local anisotropies, the actual molecular anisotropy values 
are different in different spin eigenstates. This may be rationalized 
from the fact that the energy gaps between the ground and the excited 
states are large as a consequence of which the spin correlations in these 
states are very different. We also note that in all cases, from the 
eigenvectors of the $\mathcal{D}^{(M)}$ matrix, we find that the 
choice of our laboratory frame is very close to the principal 
axis of the molecular system. 

We also examined the role of magnetic orientations 
of the core $Mn(IV)$ ions (s=3/2) and the crown 
$Mn(III)$ ions (s=2) in determining the molecular anisotropies 
by fixing the single ion orientation of the crown $Mn(III)$ ions 
at 0$^\circ$ and rotating only the orientation of the core $Mn(IV)$ 
ions systematically. The variation of $D_M$ for the S=10 ground 
state as a function of rotation of the local anisotropies of the 
core $Mn(IV)$ ions is shown in Fig. \ref{mn12Dplot}. We find that the 
molecular anisotropy is not sensitive to the local orientations 
of the core $Mn(IV)$ ions while the orientation of the 
crown $Mn(III)$ ions control the variation of the molecular 
magnetic anisotropy in $Mn_{12}Ac$. It should be noted that, in 
the case of $Mn_{12}Ac$, $E_M$ vanishes by virtue of the $D_{2d}$ 
point group symmetry to which the molecule belongs.

\begin{figure}
\includegraphics[height=6cm]{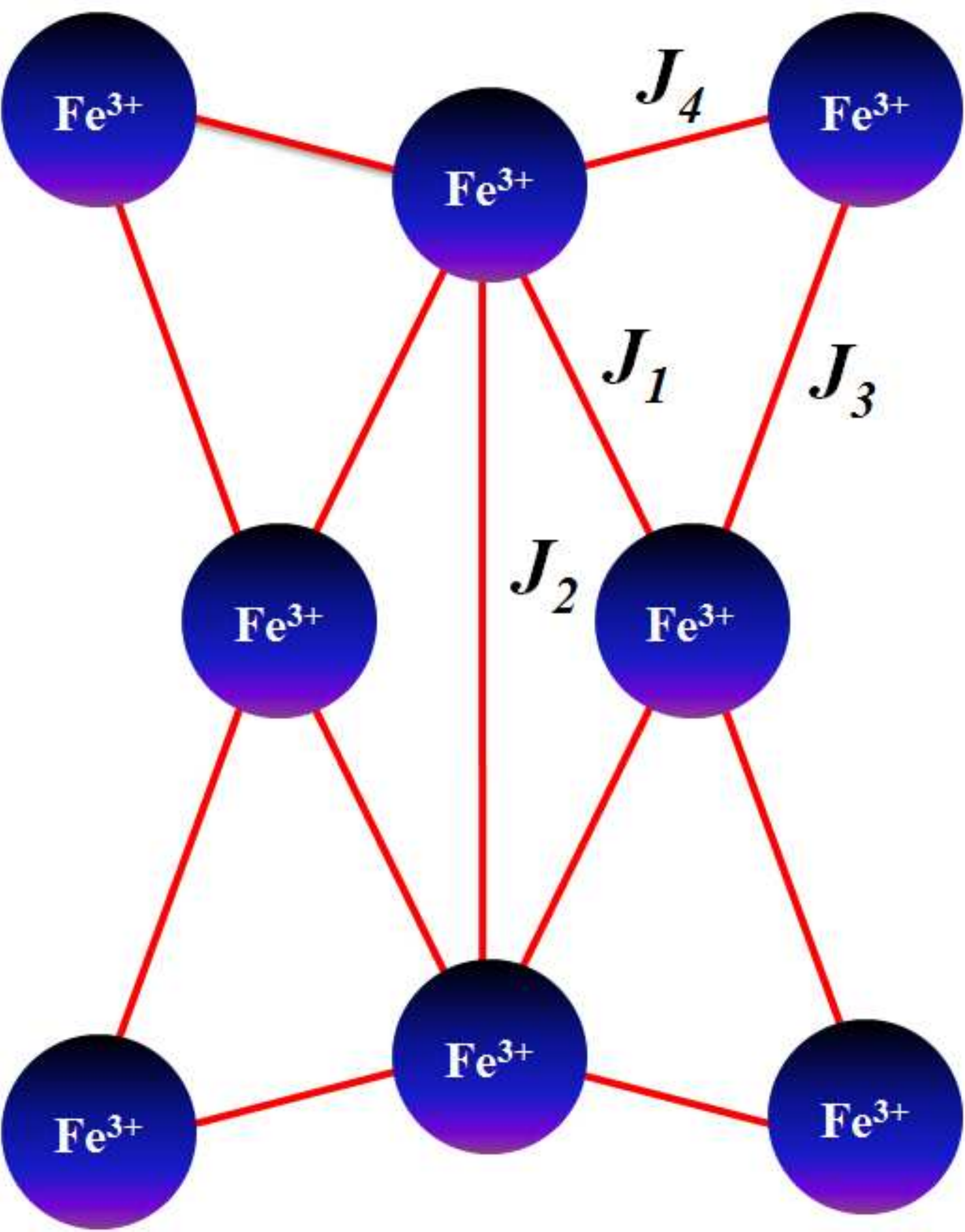}
\caption{\label{fe8exchange} Schematic of exchange 
interactions in $Fe_8$ SMM. J’s are the strength of superexchange 
interaction with $J_1$=150 K, $J_2$=25 K, $J_3$=30 K, $J_4$=50 K [9].}
\end{figure}

\subsection{\label{sec:Fe8_res}Magnetic anisotropy in $Fe_8$ SMM}
\begin{figure}
\includegraphics[height=4.5cm]{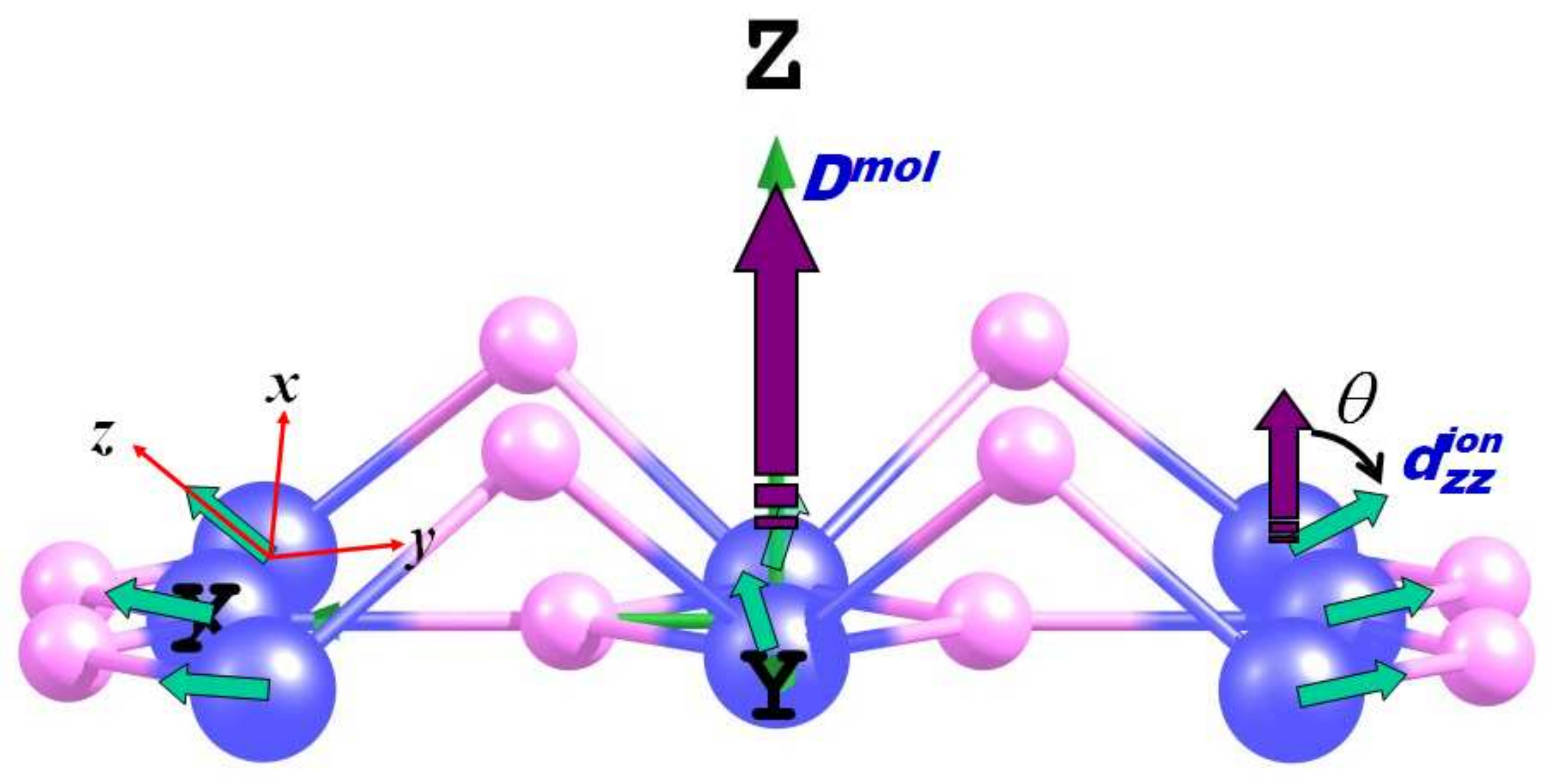}
\caption{\label{fe8scheme2} Schematic diagram showing the 
directions of local anisotropy in $Fe_8$. The $z$-component of 
the single-ion anisotropies of all the $Fe(III)$ ions are inclined 
at an angle $\theta$ to the laboratory $Z$ axis (Scheme 5).}
\end{figure}

\begin{figure}
\includegraphics[height=6cm]{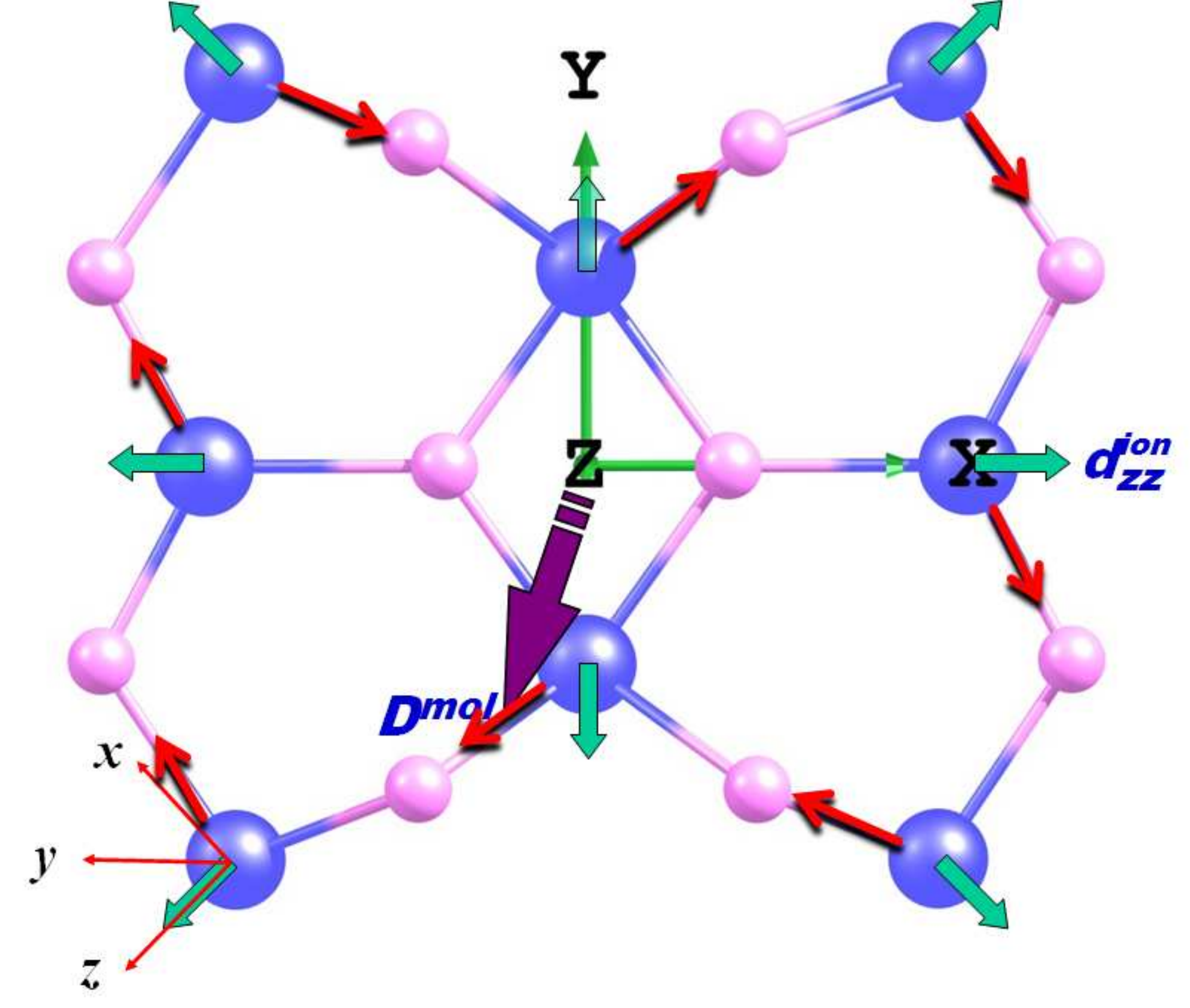}
\caption{\label{fe8scheme3} Schematic diagram showing the 
directions of local anisotropy in $Fe_8$. The $z$-component of 
the single-ion anisotropies of all the $Fe(III)$ ions are directed 
along the plane perpendicular to the laboratory $Z$ axis (Scheme 6). 
The arrows indicate the $Fe$-$O$ bonds on which the chosen local $x$-axis
has maximum projection.}
\end{figure}

\begin{figure}
\includegraphics[height=7cm]{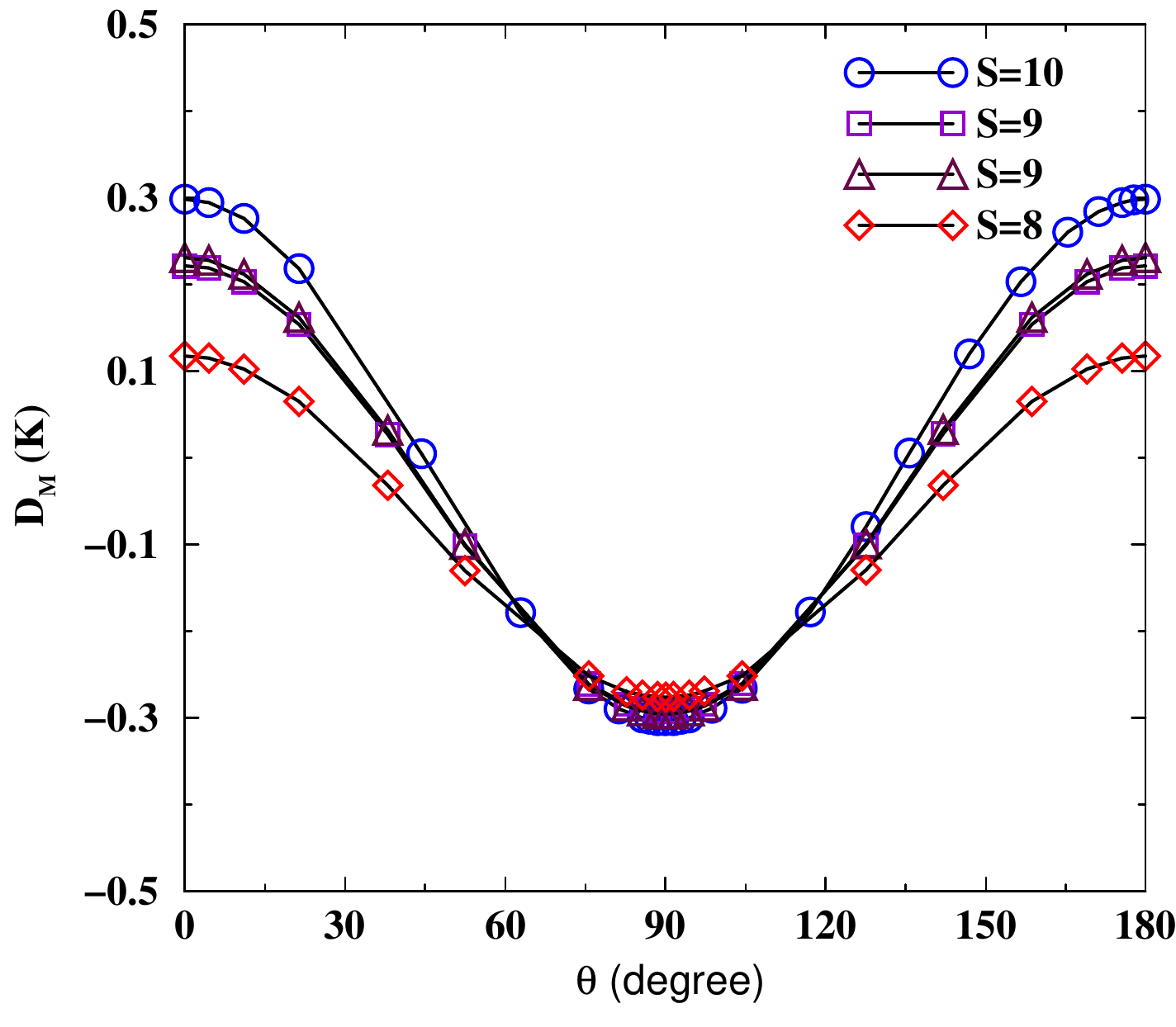}
\caption{\label{fe8Dplot} Variation of $D_M$ in $Fe_8$ 
cluster as a function of $\theta$, the angle the $z$-component 
of local anisotropy of $Fe(III)$ ions makes with the laboratory 
$Z$-axis for scheme 5.}
\end{figure}

\begin{figure}
\includegraphics[height=7cm]{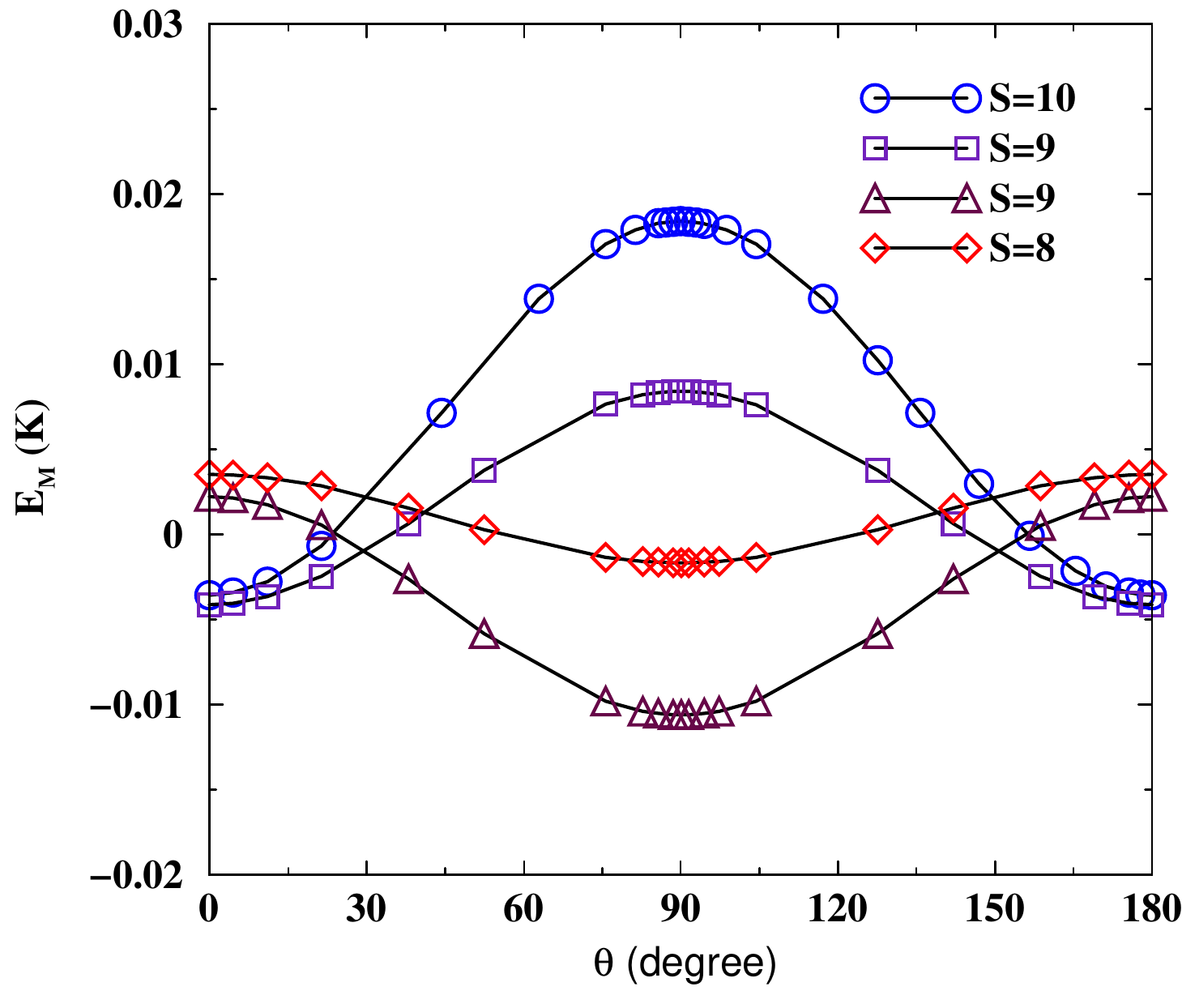}
\caption{\label{fe8Eplot} Variation of $E_M$ in $Fe_8$ 
cluster as a function of $\theta$, the angle the z-component of 
local anisotropy of $Fe(III)$ ions makes with the laboratory 
$Z$-axis for scheme 5.}
\end{figure}

We have also computed the values of molecular anisotropy for the 
$Fe_8$ molecular magnet. The unperturbed Hamiltonian in Eq. \ref{eqn3} is 
exactly solved using exchange parameters, $J_1$=150 K, $J_2$=25 K, 
$J_3$=30 K, $J_4$=50 K (Fig. \ref{fe8exchange}) \cite{raghu1}. 
The ground state of the system corresponds to total spin S=10 with a 
S=9 state at 13.56 K, a S=9 state at 27.28 K and a S=8 state at 
28.33 K above the ground state. 
To calculate the magnetic anisotropy of $Fe_8$, we have taken 
the single ion axial and rhombic anisotropy values for $Fe(III)$ 
centers to be 1.96 K and 0.008 K respectively. Using these, we have 
computed the molecular anisotropy values for three schemes  
(Schemes 4, 5 and 6) shown in Fig. \ref{fe8scheme1}, \ref{fe8scheme2} 
and \ref{fe8scheme3}. Scheme 4 corresponds to the 
case wherein the single-ion anisotropy of all the $Fe(III)$ 
ions are pointed along the laboratory $Z$ direction. In scheme 5, 
the anisotropies of the $Fe(III)$ ions are inclined at an angle 
$\theta$ to the laboratory $Z$-axis (refer Fig. \ref{fe8scheme2}). 
In Scheme 6, we have restricted the $z$-component of single-ion anisotropy 
to the laboratory $X$-$Y$ plane. We have studied the variation in 
the molecular anisotropies as a function of orientation of 
local anisotropy in these schemes for ground and the excited 
eigenstates (Fig. \ref{fe8Dplot}). We show in Table \ref{table2}, 
the $D_M$ values for the ground and the excited spin states of 
the molecule for schemes 4, 5 and 6. We note that when the local 
anisotropies are systematically varied, there is a very large variation 
in the molecular anisotropy as a function of the local orientation 
(Fig. \ref{fe8Dplot}), similar to $Mn_{12}Ac$, in all the eigenstates 
that we have studied. We also note that given the orientations 
of local anisotropies, the actual molecular anisotropy values are 
not very different in different spin eigenstates, since the 
energy gaps between the ground and the excited states are small 
and since the spin correlations in these states are not significantly 
different. The orientation of the local anisotropy centers for which 
we get the best agreement with experiments ($D_M$ = -0.28 K) 
corresponds to $\theta \sim 99^\circ$ \cite{accorsi1,heerdt1}. 
As with $Mn_{12}Ac$, we find that the laboratory frame we 
have chosen is very close to molecular axis in all the cases. 
In case of $Fe_8$ cluster, the $D_2$ symmetry commutes with the 
Hamiltonian in Eq. \ref{eqn2} and allows for a non-zero $E_M$ term. The 
variation of $E_M$ as a function of $\theta$ is shown in Fig. 
\ref{fe8Eplot}, the value of $E_M$ for which $D_M$ has the 
best fit is 0.017 K compared to the experimental estimate of 
0.046 K obtained from High-frequency EPR measurements \cite{barra2}. 

\begin{table}
\caption{\label{table2} $D_M$ values of ground and excited 
states of $Fe_8$ under schemes 4, 5 and 6 in K. For scheme 5, 
we have presented the $D_M$ values only for $\theta$=98.78$^\circ$ 
for which the $D_M$ value of the ground state matches with the 
experimentally observed value.}
~~\\
\begin{tabular}{|c|c|c|c|}\hline
\multirow{2}{*}{\bf State} & \multicolumn{3}{c|}{$\mathbf{D_M}$ {\bf (K)}}\\\cline{2-4}
      & {\bf Scheme 4}  & {\bf Scheme 5}  & {\bf Scheme 6}\\\hline
Ground state (S=10) & 0.6030 & -0.2892 & -0.3034 \\\hline
First excited state & \multirow{2}{*} {0.5821} & \multirow{2}{*} {-0.2783} & \multirow{2}{*} {-0.2923} \\
(S=9) $E_g$=13.56 K &  &  &    \\\hline
Second excited state & \multirow{2}{*} {0.5877} & \multirow{2}{*} {-0.2797} & \multirow{2}{*} {-0.2952} \\
(S=9) $E_g$=27.28 K &  &  &    \\\hline
Third excited state & \multirow{2}{*} {0.5503} & \multirow{2}{*} {-0.2694} & \multirow{2}{*} {-0.2758} \\
(S=8) $E_g$=28.33 K &  &  &    \\\hline
\end{tabular}
\end{table}

\section{\label{sec:conclusion}Conclusions}
In this paper we presented a general method to calculate the 
molecular magnetic anisotropy parameters, $D_M$ and $E_M$ for 
single molecule magnets in a chosen eigenstate of the exchange 
Hamiltonian. Since, anisotropy is generally weak in SMMs compared 
to exchange interaction, we treat the anisotropy Hamiltonian as 
a perturbation over the exchange Hamiltonian. Calculation of 
$D_M$ and $E_M$ values assuming only dipolar interactions 
between the transition metal ions give negligible values of 
molecular magnetic anisotropy compared to the experimental 
values. Therefore, we focus on the molecular anisotropy from the 
single-ion anisotropies of the individual transition metal 
centers in the SMM. The single-ion anisotropy has relativistic 
origin, (spin-orbit interactions generally dominate over 
dipolar interactions between unpaired electrons in case of 
transition metal ions) and are hence short ranged with inverse 
cube dependence on distance. Therefore, we neglect interaction 
between spin moment on one ion with the orbital moment on another. 
This approximation simplifies the perturbation Hamiltonian. 
The molecular anisotropies are computed from the single-ion 
anisotropies, using first order perturbation theory for different 
spin states of the SMMs. We have computed the molecular magnetic 
anisotropy parameters of $Mn_{12}Ac$ and $Fe_8$ SMMs in various 
eigenstates of different total spin. We also studied the variation 
of molecular anisotropy by rotating the local anisotropy of the 
metal ions. In case of $Mn_{12}Ac$, we find that the molecular 
anisotropy changes drastically with the local anisotropy direction. 
The $D_M$ value is different in ground and excited states we have 
computed, owing to large difference in spin-spin correlation values. 
The molecular anisotropy of $Mn_{12}Ac$ does not change significantly 
with the orientation of the local anisotropy of the core Mn(IV) 
ions. In the case of $Fe_8$ cluster also, we find that the 
molecular anisotropy parameters depend strongly on the 
orientation of the local anisotropy. $D_M$ value is not very different 
in ground and excited states probably due to small energy gaps 
which implies similar spin-spin correlations. In case of $Mn_{12}Ac$, 
the first order rhombic anisotropy term is zero due to the $D_{2d}$ 
symmetry of the molecule while it is non-zero in $Fe_8$. The second 
order rhombic anisotropy terms commute with the molecular symmetry 
of the $Mn_{12}Ac$ cluster and cause tunneling between the states 
on either side of the double potential well. Our method can 
also be extended to the calculation of these anisotropy constants.

\begin{acknowledgments}
The authors thank Indo-French Centre for Promotion of 
Advanced Research (IFCPAR) under the project 3108-3 and 
Indo-Swedish Research Council under the project VT-SK001 for the 
financial support. 
\end{acknowledgments}

\bibliography{Raghunathan_anisotropy}

\begin{thebibliography}{23}
\expandafter\ifx\csname natexlab\endcsname\relax\def\natexlab#1{#1}\fi
\expandafter\ifx\csname bibnamefont\endcsname\relax
  \def\bibnamefont#1{#1}\fi
\expandafter\ifx\csname bibfnamefont\endcsname\relax
  \def\bibfnamefont#1{#1}\fi
\expandafter\ifx\csname citenamefont\endcsname\relax
  \def\citenamefont#1{#1}\fi
\expandafter\ifx\csname url\endcsname\relax
  \def\url#1{\texttt{#1}}\fi
\expandafter\ifx\csname urlprefix\endcsname\relax\def\urlprefix{URL }\fi
\providecommand{\bibinfo}[2]{#2}
\providecommand{\eprint}[2][]{\url{#2}}

\bibitem[{\citenamefont{Thomas et~al.}(1996)\citenamefont{Thomas, Lionti,
  Ballou, Gatteschi, Sessoli, and Barbara}}]{thomas1}
\bibinfo{author}{\bibfnamefont{L.}~\bibnamefont{Thomas}},
  \bibinfo{author}{\bibfnamefont{F.}~\bibnamefont{Lionti}},
  \bibinfo{author}{\bibfnamefont{R.}~\bibnamefont{Ballou}},
  \bibinfo{author}{\bibfnamefont{D.}~\bibnamefont{Gatteschi}},
  \bibinfo{author}{\bibfnamefont{R.}~\bibnamefont{Sessoli}}, \bibnamefont{and}
  \bibinfo{author}{\bibfnamefont{B.}~\bibnamefont{Barbara}},
  \bibinfo{journal}{Nature} \textbf{\bibinfo{volume}{383}},
  \bibinfo{pages}{145} (\bibinfo{year}{1996}).

\bibitem[{\citenamefont{Linert and Verdaguer}(2003)}]{linert1}
\bibinfo{author}{\bibfnamefont{W.}~\bibnamefont{Linert}} \bibnamefont{and}
  \bibinfo{author}{\bibfnamefont{M.}~\bibnamefont{Verdaguer}},
  \emph{\bibinfo{title}{Molecular Magnets: Recent Highlights}}
  (\bibinfo{publisher}{Springer-Verlag, New York}, \bibinfo{year}{2003}).

\bibitem[{\citenamefont{Miller and Drillon}(2003)}]{miller1}
\bibinfo{author}{\bibfnamefont{J.~S.} \bibnamefont{Miller}} \bibnamefont{and}
  \bibinfo{author}{\bibfnamefont{M.}~\bibnamefont{Drillon}},
  \emph{\bibinfo{title}{Magnetism: Molecules to Materials IV}}
  (\bibinfo{publisher}{Wiley-VCH Verlag GMBH}, \bibinfo{year}{2003}).

\bibitem[{\citenamefont{Shoji et~al.}(2006)\citenamefont{Shoji, Koizumi,
  Hamamoto, Kitagawa, Yamanaka, Okumura, and Yamaguchi}}]{shoji1}
\bibinfo{author}{\bibfnamefont{M.}~\bibnamefont{Shoji}},
  \bibinfo{author}{\bibfnamefont{K.}~\bibnamefont{Koizumi}},
  \bibinfo{author}{\bibfnamefont{T.}~\bibnamefont{Hamamoto}},
  \bibinfo{author}{\bibfnamefont{Y.}~\bibnamefont{Kitagawa}},
  \bibinfo{author}{\bibfnamefont{S.}~\bibnamefont{Yamanaka}},
  \bibinfo{author}{\bibfnamefont{M.}~\bibnamefont{Okumura}}, \bibnamefont{and}
  \bibinfo{author}{\bibfnamefont{K.}~\bibnamefont{Yamaguchi}},
  \bibinfo{journal}{Chem. Phys. Lett.} \textbf{\bibinfo{volume}{421}},
  \bibinfo{pages}{483} (\bibinfo{year}{2006}).

\bibitem[{\citenamefont{Wieghardt et~al.}(1984)\citenamefont{Wieghardt, Pohl,
  Jibril, and Huttner}}]{wieghardt1}
\bibinfo{author}{\bibfnamefont{K.}~\bibnamefont{Wieghardt}},
  \bibinfo{author}{\bibfnamefont{K.}~\bibnamefont{Pohl}},
  \bibinfo{author}{\bibfnamefont{I.}~\bibnamefont{Jibril}}, \bibnamefont{and}
  \bibinfo{author}{\bibfnamefont{G.}~\bibnamefont{Huttner}},
  \bibinfo{journal}{Angew. Chem. Int. Ed. Engl.} \textbf{\bibinfo{volume}{24}},
  \bibinfo{pages}{77} (\bibinfo{year}{1984}).

\bibitem[{\citenamefont{Barco et~al.}(1999)\citenamefont{Barco, Vernier,
  Hernandez, Tejada, Chudnovski, Molins, and Bellessa}}]{barco1}
\bibinfo{author}{\bibfnamefont{E.~D.} \bibnamefont{Barco}},
  \bibinfo{author}{\bibfnamefont{N.}~\bibnamefont{Vernier}},
  \bibinfo{author}{\bibfnamefont{J.~M.} \bibnamefont{Hernandez}},
  \bibinfo{author}{\bibfnamefont{J.}~\bibnamefont{Tejada}},
  \bibinfo{author}{\bibfnamefont{E.~M.} \bibnamefont{Chudnovski}},
  \bibinfo{author}{\bibfnamefont{E.}~\bibnamefont{Molins}}, \bibnamefont{and}
  \bibinfo{author}{\bibfnamefont{G.}~\bibnamefont{Bellessa}},
  \bibinfo{journal}{Euro. Phys. Lett.} \textbf{\bibinfo{volume}{47}},
  \bibinfo{pages}{722} (\bibinfo{year}{1999}).

\bibitem[{\citenamefont{Sessoli et~al.}(1993)\citenamefont{Sessoli, Gatteschi,
  Canneschi, and Novak}}]{sessoli1}
\bibinfo{author}{\bibfnamefont{R.}~\bibnamefont{Sessoli}},
  \bibinfo{author}{\bibfnamefont{D.}~\bibnamefont{Gatteschi}},
  \bibinfo{author}{\bibfnamefont{A.}~\bibnamefont{Canneschi}},
  \bibnamefont{and} \bibinfo{author}{\bibfnamefont{M.~A.} \bibnamefont{Novak}},
  \bibinfo{journal}{Nature} \textbf{\bibinfo{volume}{365}},
  \bibinfo{pages}{141} (\bibinfo{year}{1993}).

\bibitem[{\citenamefont{Sahoo et~al.}(2008)\citenamefont{Sahoo, Rajamani,
  Ramasesha, and Sen}}]{sahoo1}
\bibinfo{author}{\bibfnamefont{S.}~\bibnamefont{Sahoo}},
  \bibinfo{author}{\bibfnamefont{R.}~\bibnamefont{Rajamani}},
  \bibinfo{author}{\bibfnamefont{S.}~\bibnamefont{Ramasesha}},
  \bibnamefont{and} \bibinfo{author}{\bibfnamefont{D.}~\bibnamefont{Sen}},
  \bibinfo{journal}{arXiv:0804.1319}  (\bibinfo{year}{2008}).

\bibitem[{\citenamefont{Raghu et~al.}(2001)\citenamefont{Raghu, Rudra, Sen, and
  Ramasesha}}]{raghu1}
\bibinfo{author}{\bibfnamefont{C.}~\bibnamefont{Raghu}},
  \bibinfo{author}{\bibfnamefont{I.}~\bibnamefont{Rudra}},
  \bibinfo{author}{\bibfnamefont{D.}~\bibnamefont{Sen}}, \bibnamefont{and}
  \bibinfo{author}{\bibfnamefont{S.}~\bibnamefont{Ramasesha}},
  \bibinfo{journal}{Phys. Rev. B} \textbf{\bibinfo{volume}{64}},
  \bibinfo{pages}{064419} (\bibinfo{year}{2001}).

\bibitem[{\citenamefont{Sinnecker and Neese}(2006)}]{sinnecker1}
\bibinfo{author}{\bibfnamefont{S.}~\bibnamefont{Sinnecker}} \bibnamefont{and}
  \bibinfo{author}{\bibfnamefont{F.}~\bibnamefont{Neese}}, \bibinfo{journal}{J.
  Phys. Chem A} \textbf{\bibinfo{volume}{110}}, \bibinfo{pages}{12267}
  (\bibinfo{year}{2006}).

\bibitem[{\citenamefont{Baruah and Pederson}(2002)}]{baruah1}
\bibinfo{author}{\bibfnamefont{T.}~\bibnamefont{Baruah}} \bibnamefont{and}
  \bibinfo{author}{\bibfnamefont{M.~R.} \bibnamefont{Pederson}},
  \bibinfo{journal}{Chem. Phys. Lett.} \textbf{\bibinfo{volume}{360}},
  \bibinfo{pages}{144} (\bibinfo{year}{2002}).

\bibitem[{\citenamefont{Arino et~al.}(2005)\citenamefont{Arino, Baruah, and
  Pederson}}]{arino1}
\bibinfo{author}{\bibfnamefont{J.~R.} \bibnamefont{Arino}},
  \bibinfo{author}{\bibfnamefont{T.}~\bibnamefont{Baruah}}, \bibnamefont{and}
  \bibinfo{author}{\bibfnamefont{M.~R.} \bibnamefont{Pederson}},
  \bibinfo{journal}{J. Chem. Phys.} \textbf{\bibinfo{volume}{123}},
  \bibinfo{pages}{044303} (\bibinfo{year}{2005}).

\bibitem[{\citenamefont{Oudar and Zyss}(1982)}]{oudar1}
\bibinfo{author}{\bibfnamefont{J.~L.} \bibnamefont{Oudar}} \bibnamefont{and}
  \bibinfo{author}{\bibfnamefont{J.}~\bibnamefont{Zyss}},
  \bibinfo{journal}{Phys. Rev. A} \textbf{\bibinfo{volume}{26}},
  \bibinfo{pages}{2016} (\bibinfo{year}{1982}).

\bibitem[{\citenamefont{Bencini et~al.}(1984)\citenamefont{Bencini, Benelli,
  and Gatteschi}}]{bencini1}
\bibinfo{author}{\bibfnamefont{A.}~\bibnamefont{Bencini}},
  \bibinfo{author}{\bibfnamefont{C.}~\bibnamefont{Benelli}}, \bibnamefont{and}
  \bibinfo{author}{\bibfnamefont{D.}~\bibnamefont{Gatteschi}},
  \bibinfo{journal}{Coord. Chem. Rev.} \textbf{\bibinfo{volume}{60}},
  \bibinfo{pages}{131} (\bibinfo{year}{1984}).

\bibitem[{\citenamefont{Bencini and Gatteschi}(1990)}]{bencini2}
\bibinfo{author}{\bibfnamefont{A.}~\bibnamefont{Bencini}} \bibnamefont{and}
  \bibinfo{author}{\bibfnamefont{D.}~\bibnamefont{Gatteschi}},
  \emph{\bibinfo{title}{EPR of Exchange Coupled Systems}}
  (\bibinfo{publisher}{Springer, Berlin}, \bibinfo{year}{1990}).

\bibitem[{\citenamefont{Carrington and McLachlan}(1967)}]{carrington1}
\bibinfo{author}{\bibfnamefont{A.}~\bibnamefont{Carrington}} \bibnamefont{and}
  \bibinfo{author}{\bibfnamefont{A.}~\bibnamefont{McLachlan}},
  \emph{\bibinfo{title}{Introduction to Magnetic Ressonance: with applications
  to chemistry and chemical physics}} (\bibinfo{publisher}{Harrer and Row
  Publishers, New York}, \bibinfo{year}{1967}).

\bibitem[{\citenamefont{Ramasesha and Soos}(1984)}]{ramasesha1}
\bibinfo{author}{\bibfnamefont{S.}~\bibnamefont{Ramasesha}} \bibnamefont{and}
  \bibinfo{author}{\bibfnamefont{Z.~G.} \bibnamefont{Soos}},
  \bibinfo{journal}{Chem. Phys.} \textbf{\bibinfo{volume}{91}},
  \bibinfo{pages}{35} (\bibinfo{year}{1984}).

\bibitem[{\citenamefont{Barra et~al.}(2000)\citenamefont{Barra, Gatteschi, and
  Sessoli}}]{barra1}
\bibinfo{author}{\bibfnamefont{A.~L.} \bibnamefont{Barra}},
  \bibinfo{author}{\bibfnamefont{D.}~\bibnamefont{Gatteschi}},
  \bibnamefont{and} \bibinfo{author}{\bibfnamefont{R.}~\bibnamefont{Sessoli}},
  \bibinfo{journal}{Chem. Eur. J.} \textbf{\bibinfo{volume}{6}},
  \bibinfo{pages}{1608} (\bibinfo{year}{2000}).

\bibitem[{\citenamefont{Barra et~al.}(1996)\citenamefont{Barra, Debrunner,
  Gatteschi, Schulz, and Sessoli}}]{barra2}
\bibinfo{author}{\bibfnamefont{A.-L.} \bibnamefont{Barra}},
  \bibinfo{author}{\bibfnamefont{P.}~\bibnamefont{Debrunner}},
  \bibinfo{author}{\bibfnamefont{D.}~\bibnamefont{Gatteschi}},
  \bibinfo{author}{\bibfnamefont{C.~E.} \bibnamefont{Schulz}},
  \bibnamefont{and} \bibinfo{author}{\bibfnamefont{R.}~\bibnamefont{Sessoli}},
  \bibinfo{journal}{Europhys. Lett.} \textbf{\bibinfo{volume}{35}},
  \bibinfo{pages}{133} (\bibinfo{year}{1996}).

\bibitem[{\citenamefont{Cornia et~al.}(2002)\citenamefont{Cornia, Sessoli,
  Sorace, Gatteschi, Barra, and Daiguebonne}}]{cornia1}
\bibinfo{author}{\bibfnamefont{A.}~\bibnamefont{Cornia}},
  \bibinfo{author}{\bibfnamefont{R.}~\bibnamefont{Sessoli}},
  \bibinfo{author}{\bibfnamefont{L.}~\bibnamefont{Sorace}},
  \bibinfo{author}{\bibfnamefont{D.}~\bibnamefont{Gatteschi}},
  \bibinfo{author}{\bibfnamefont{A.~L.} \bibnamefont{Barra}}, \bibnamefont{and}
  \bibinfo{author}{\bibfnamefont{C.}~\bibnamefont{Daiguebonne}},
  \bibinfo{journal}{Phys. Rev. Lett.} \textbf{\bibinfo{volume}{89}},
  \bibinfo{pages}{257201} (\bibinfo{year}{2002}).

\bibitem[{\citenamefont{Barra et~al.}(1997)\citenamefont{Barra, Gatteschi,
  Sessoli, Abbati, Cornia, Fabretti, and Uytterhoeven}}]{barra3}
\bibinfo{author}{\bibfnamefont{A.~L.} \bibnamefont{Barra}},
  \bibinfo{author}{\bibfnamefont{D.}~\bibnamefont{Gatteschi}},
  \bibinfo{author}{\bibfnamefont{R.}~\bibnamefont{Sessoli}},
  \bibinfo{author}{\bibfnamefont{G.~L.} \bibnamefont{Abbati}},
  \bibinfo{author}{\bibfnamefont{A.}~\bibnamefont{Cornia}},
  \bibinfo{author}{\bibfnamefont{A.~C.} \bibnamefont{Fabretti}},
  \bibnamefont{and} \bibinfo{author}{\bibfnamefont{M.~G.}
  \bibnamefont{Uytterhoeven}}, \bibinfo{journal}{Angew. Chem. Int. Ed. Engl.}
  \textbf{\bibinfo{volume}{36}}, \bibinfo{pages}{2329} (\bibinfo{year}{1997}).

\bibitem[{\citenamefont{Accorsi et~al.}(2006)\citenamefont{Accorsi, Barra,
  Caneschi, Chastanet, Cornia, Fabretti, Gatteschi, Mortalo, Olivieri, Parenti
  et~al.}}]{accorsi1}
\bibinfo{author}{\bibfnamefont{S.}~\bibnamefont{Accorsi}},
  \bibinfo{author}{\bibfnamefont{A.-L.} \bibnamefont{Barra}},
  \bibinfo{author}{\bibfnamefont{A.}~\bibnamefont{Caneschi}},
  \bibinfo{author}{\bibfnamefont{G.}~\bibnamefont{Chastanet}},
  \bibinfo{author}{\bibfnamefont{A.}~\bibnamefont{Cornia}},
  \bibinfo{author}{\bibfnamefont{A.}~\bibnamefont{Fabretti}},
  \bibinfo{author}{\bibfnamefont{D.}~\bibnamefont{Gatteschi}},
  \bibinfo{author}{\bibfnamefont{C.}~\bibnamefont{Mortalo}},
  \bibinfo{author}{\bibfnamefont{E.}~\bibnamefont{Olivieri}},
  \bibinfo{author}{\bibfnamefont{F.}~\bibnamefont{Parenti}},
  \bibnamefont{et~al.}, \bibinfo{journal}{J. Am. Chem. Soc.}
  \textbf{\bibinfo{volume}{128}}, \bibinfo{pages}{4742} (\bibinfo{year}{2006}).

\bibitem[{\citenamefont{t.~Heerdt et~al.}(2006)\citenamefont{t.~Heerdt, Stefan,
  Goovaerts, Canneschi, and Cornia}}]{heerdt1}
\bibinfo{author}{\bibfnamefont{P.}~\bibnamefont{t.~Heerdt}},
  \bibinfo{author}{\bibfnamefont{M.}~\bibnamefont{Stefan}},
  \bibinfo{author}{\bibfnamefont{E.}~\bibnamefont{Goovaerts}},
  \bibinfo{author}{\bibfnamefont{A.}~\bibnamefont{Canneschi}},
  \bibnamefont{and} \bibinfo{author}{\bibfnamefont{A.}~\bibnamefont{Cornia}},
  \bibinfo{journal}{J. Mag. Res.} \textbf{\bibinfo{volume}{179}},
  \bibinfo{pages}{29} (\bibinfo{year}{2006}).

\end{thebibliography}

\end{document}